\documentclass[journal]{IEEEtran}

\ifCLASSINFOpdf
\else
\fi

\usepackage{algorithm}
\usepackage{algorithmic}
\usepackage[fleqn]{amsmath}
\usepackage{multirow}
\usepackage{graphicx}
\usepackage{url}
\usepackage[shortlabels]{enumitem}
\usepackage{color}
\definecolor{blue}{rgb}{0.0, 0.46, 0.37}

\begin{document}
	\title{Phase asymmetry ultrasound despeckling with fractional anisotropic diffusion and total variation}

\author{Kunqiang~Mei, Bin Hu, Baowei Fei, and Binjie~Qin*,~\IEEEmembership{Member,~IEEE}
\thanks{Manuscript received April 14, 2018; accepted October 27, 2018; This work was supported by the National Natural Science Foundation of China (61271320), Medical Engineering Cross Fund of Shanghai Jiao Tong University (YG2014MS29), and Translational Medicine Cross Fund of Shanghai Jiao Tong University (ZH2018ZDA19).}
\thanks{Kunqiang Mei and Binjie Qin are with the School of Biomedical Engineering, Shanghai Jiao Tong University, Shanghai, 200240, China. E-mail: bjqin@sjtu.edu.cn.}% <-this % stops a space
\thanks{Bin Hu is with Department of Ultrasound in Medicine, Shanghai Jiao Tong University Affiliated Sixth People’s Hospital, Shanghai Institute of Ultrasound in Medicine, Shanghai, 200233, China.}
%niuniuhu1213@qq.com
\thanks{Baowei Fei is with Department of Bioengineering, Erik Jonsson School of Engineering and Computer Science, University of Texas at Dallas, Richardson, TX 75080, USA.}}

\markboth{Journal of \LaTeX\ Class Files,~Vol.~14, No.~8, August~2019}%
{Shell \MakeLowercase{\textit{et al.}}: Bare Demo of IEEEtran.cls for IEEE Journals}

\maketitle

\begin{abstract}
We propose an ultrasound speckle filtering method for not only preserving various edge features but also filtering tissue-dependent complex speckle noises in ultrasound images. The key idea is to detect these various edges using a phase congruence-based edge significance measure called phase asymmetry (PAS), which is invariant to the intensity amplitude of edges and takes 0 in non-edge smooth regions and 1 at the idea step edge, while also taking intermediate values at slowly varying ramp edges. By leveraging the PAS metric in designing weighting coefficients to maintain a balance between fractional-order anisotropic diffusion and total variation (TV) filters in TV cost function, we propose a new fractional TV framework to not only achieve the best despeckling performance with ramp edge preservation but also reduce the staircase effect produced by integral-order filters. Then, we exploit the PAS metric in designing a new fractional-order diffusion coefficient to properly preserve low-contrast edges in diffusion filtering. Finally, different from fixed fractional-order diffusion filters, an adaptive fractional order is introduced based on the PAS metric to enhance various weak edges in the spatially transitional areas between objects. The proposed fractional TV model is minimized using the gradient descent method to obtain the final denoised image. The experimental results and real application of ultrasound breast image segmentation show that the proposed method outperforms other state-of-the-art ultrasound despeckling filters for both speckle reduction and feature preservation in terms of visual evaluation and quantitative indices. The best scores on feature similarity indices have achieved 0.867, 0.844 and 0.834 under three different levels of noise, while the best breast ultrasound segmentation accuracy in terms of the mean and median dice similarity coefficient are 96.25\% and 96.15\%, respectively.  
\end{abstract}

\begin{IEEEkeywords}
ultrasound despeckling, speckle noise, fractional-order diffusion filter, fractional-order TV filter, edge detection, phase congruency, phase asymmetry, image denoising.
\end{IEEEkeywords}
\IEEEpeerreviewmaketitle

\section{Introduction}
\IEEEPARstart{C}{urrent} advanced image denoising \cite{1}\cite{2} and image enhancement algorithms \cite{3}\cite{4} are developed to solve the difficult and urgent open problems of feature preservation in the removal of complex signal-dependent noises for challenging real applications. In these recent research developments, feature-preserving ultrasound despeckling \cite{5}\cite{6} is most desired in clinical applications due to the ubiquity of the ultrasound imaging modality given its noninvasiveness, low cost and convenience; however, its quality is relatively poor compared with other medical imaging modalities. The main reason for quality degradation in ultrasound images is the presence of an inherent imaging artefact called speckle, which results from constructive and destructive coherent interferences of backscattered echoes from the scatterers \cite{5}\cite{6}\cite{7}. Speckle is commonly interpreted as a locally correlated noise that reduces image contrast and conceals fine feature details \cite{8}, causing negative effects on medical diagnosis and reduction of the accuracy of subsequent image processing such as segmentation and registration \cite{7}. Furthermore, extracting coherent feature patterns from the noisy ultrasound signals is necessary for super-resolution ultrasound microvessel imaging \cite{9} and 3D reconstruction from a series of 2D freehand ultrasound images \cite{10}. Therefore, it is very important to remove speckle noise with satisfactory feature preservation for accurate diagnosis and analysis in many applications. 

However, feature-preserving speckle reduction is a challenging task, since speckle noise is known to be tissue-dependent and it manifests itself in the form of multiplicative noise, which means that the intensity of speckle can change sharply and the intensity of variance of speckle is comparable to or even larger than that of the features\cite{11}. Therefore, simply employing intensity-based gradient information cannot accurately distinguish edges from speckle noise, especially for low-contrast edges. Failing to preserve various edges will damage other features of structures (or shapes) whose boundaries are composed of the edges. Existing edge-preserving image processing techniques \cite{13}\cite{14} are likely to damage some low-contrast features, since they regard some low-contrast edges as speckle noise and remove these edges after noise removal by exploring intensity-based gradient information.

Various speckle reduction filters are proposed to solve the abovementioned challenges, including local adaptive filters, non-local means (NLM) filters and diffusion filters. The local adaptive filters such as Frost \cite{14} filters and bilateral filters \cite{15} rectify a pixel by averaging its neighbouring pixels. In their experimental comparison, Chen \textit{et al.} \cite{16} indicated that the bilateral filtering is not as good as the other methods such as diffusion filters for the edge preservation in ultrasound images. Moreover, the squeeze box filter (SBF) \cite{17} rectifies only local extrema at each iteration by replacing them with the local mean. However, these local adaptive filters are sensitive to the shape and size of local windows. The NLM algorithms assume that natural images contain many similar features. NLM algorithms group similar features from  different image patches and remove noise by a weighted average of similar features. Coupe \textit{et al.} \cite{18} proposed the optimal Bayesian NLM (OBNLM) filter to process ultrasound images in non-Gaussian speckle noise circumstances. An improved OBNLM filter is proposed by Zhou \textit{et al.} \cite{19} to iteratively refine the filtering model by deducing the key probability density function according to the statistical characteristic of the speckle noise. Recently, Zhu \textit{et al.} \cite{20} developed a non-local low-rank framework (NLLRF) for ultrasound speckle reduction, which leverages a guidance image to improve the performance of patch selection. However, NLM algorithms usually mix different features into the same patch cluster in the case of a large number of features, causing some important details to become indiscernible after noise removal \cite{21}.

In regard to diffusion filtering, after Perona and Malik proposed the well-known anisotropic diffusion (AD) filter \cite{22}, both the speckle reducing anisotropic diffusion (SRAD) \cite{23} filter and the detail preserving anisotropic diffusion (DPAD) \cite{24} filter have been modified based on the AD filter. The SRAD filter added a parameter related to the noise estimate into the diffusion coefficient, while the DPAD filter adopted an improved noise estimator to improve the despeckling performance. The oriented speckle reducing anisotropic diffusion \cite{25} filter modified the diffusion coefficient with the local directional variance of the image intensity. Using an edge indicator to distinguish between sharp and ramp edges in images, Chen \textit{et al.} \cite{16} proposed to adaptively determine the diffusion coefficient for introducing isotropic diffusion in flat and ramp regions and anisotropic diffusion in sharp edges for medical images. Using the information about image gradient and grey levels of the image, a doubly degenerate diffusion model \cite{26} with robust speckle reduction performance \cite{27} is proposed to remove multiplicative noise. However, all of these diffusion filters employ the local intensity-based gradient or grey level information to identify edges, failing to detect and preserve low-contrast edges. Moreover, the gradient difference between sharp edges and ramp edges is not obvious, and therefore anisotropic diffusions based on the image gradient are prone to cause the staircase effect in the regions of ramp edges. To reduce the staircase effect, Bai and Feng \cite{28} proposed a fixed fractional-order AD (FAD) model for image denoising. Nevertheless, the fixed fractional-order diffusion filter neglects the differences among various image regions. Recently, Flores \textit{et al.} \cite{8} developed an anisotropic diffusion filter guided by the log-Gabor filters (ADLG) instead of the intensity-based gradient. However, ADLG fails to achieve satisfactory feature preservation. Based on nonlinear AD for filtering noisy coefficients in the transform domain, some wavelet \cite{29} and shearlet-based \cite{30} transform-domain filtering approaches are combined with AD to exploit the advantages of multi-resolution analysis, noise removal and edge preservation. However, these methods have a high computational cost due to the transformation and anti-transformation steps and may insert or manipulate artificial frequencies in the recovered image.

%On Detection of Faint Edges in Noisy Images 
Essentially, edge detection and image denoising depend on each other, leading to a “chicken-or-the-egg” causality dilemma in ultrasound despeckling. The abovementioned diffusion filters that fail to accurately identify edges from speckle noises cannot achieve satisfactory feature preservation after speckle filtering. As an important image feature, the edge is a basic element of other features, such as ridges, valleys, textures and boundaries of structures, such that failing to preserve the edges will render these structure features inviable \cite{31}. Therefore, robustly detecting the edge from the noisy image is half the battle of ultrasound despeckling. To solve the drawback of local methods \cite{32} that use local intensity information for edge detection, Ofir \textit{et al.} \cite{33} consider the edge detection as a search approach in a large set of feasible curves by hierarchically constructing difference filters that match the curves traced by the sought edges. However, this method has a high computational cost on large and noisy images containing long edges and has a great limitation in searching weak irregular texture boundaries \cite{2} in noisy images. Cogranne \textit{et al.} \cite{34} implement edge detection as a regression problem with a regression surface based on local image content model. Such an idea of transferring image processing into searching a regression's coefficients from the context information of a neighbourhood area is successfully and widely used in image denoising \cite{35}, image reconstruction \cite{36}, image registration \cite{37}\cite{38}, and so on. However the computation of the regression's coefficients for indicating the edges' presence is still sensitive to the sharp change of local intensity information in the presence of multiplicative noise.
 
\begin{table}
	\centering
	\fontsize{9}{14}\selectfont
	\caption{Acronyms.}
	\label{tab2}
	\begin{tabular}{lll}
		\hline
		&Acronym  &Definition\cr\hline
		&PAS  &Phase asymmetry\cr
		&PS  &Phase symmetry\cr
		&PC  &Phase congruency\cr
		&AD  &Anisotropic diffusion\cr
		&FAD &Fractional-order anisotropic diffusion\cr
		&TV &Total variation\cr
		&FTV &Fractional-order TV\cr
		&NLM &Non-local means\cr
		&SBF &Squeeze box filter\cr
		&OBNLM &Optimal Bayesian NLM\cr
		&SRAD  &Speckle reducing anisotropic diffusion\cr
		&DPAD  &Detail preserving anisotropic diffusion\cr
		&ADLG  &AD filter guided by the log-Gabor filter\cr
		&PFDTV &Phase asymmetry fractional AD and TV\cr
		&G-L   &Gr\"unwald–Letnikov\cr
		&PSNR  &Peak signal-to-noise ratio\cr
		&FSIM  &Feature similarity index\cr
		&MSSIM &Mean structural similarity index\cr
		&BUS   &Breast ultrasound\cr
		&DSC   &Dice similarity coefficient\cr
		&JS    &Jaccard similarity\cr
		&HD    &Hausdorff distance\cr
		&HM    &Hausdorff mean\cr\hline
	\end{tabular}
\end{table}

To solve the drawback of intensity-based edge detectors, some local phase-based edge detection methods \cite{39}\cite{40} were developed in ultrasound imaging. In analytically representing images using spatially varying sinusoidal waves, the local phase informs us about the location and orientation of image features, while the amplitude provides only information on their intensity. More specifically, local phase is an illumination and contrast-invariant measure of feature significance based on a model of feature perception called the local-energy model developed by Morrone \textit{et al.} \cite{41}\cite{42}. This model postulates that features are perceived at points in an image, where the Fourier components are maximally in phase. A wide range of feature types give rise to points of high-phase congruency. These include step edges, line and roof edges, and Mach bands. The first local phase-based work for successfully detecting the boundary in echocardiographic images is from Mulet-Parada and Noble \cite{43}. We then use the method introduced in \cite{44} to construct a phase congruency (PC)-based feature indicator called phase symmetry (PS) and phase asymmetry (PAS, all acronyms in this paper are listed in Table I) depending on the image feature type to be detected. Both PS and PAS are special patterns of PC \cite{44}. Specifically, PS can be used to detect symmetry features for image enhancement such as bone surface enhancement \cite{45} and vessel enhancement \cite{4}, while PAS can be used to identify edge features for ultrasound image denoising \cite{39} and segmentation \cite{40}. Zhu \textit{et al.} \cite{39} employ the phase congruency-based PAS metric to extract edge features on the phase domain for efficiently and robustly distinguishing features and speckle noise during the ultrasound despeckling process. Inspired by these works, we use the PAS to detect various types of edge features. In ultrasound images, almost all of the real edges are weak ramp edges of various slopes rather than ideal step edges. Particularly, low-contrast edges refer to the weak ramp edges with low-step amplitude. In regard to the symmetry features in images, they are also made of edges. If we preserve their edges using the PAS indicator, the symmetry features will be preserved properly.

The PAS metric provides a good measure of asymmetric image features and can effectively detect various edges from backgrounds of images\cite{39}\cite{40}\cite{45}. Specifically, the PAS metric represents the edge significance \cite{42}\cite{44} of each point and varies from $0$ to $1$, taking $0$ (no-edge significance) in ideal smooth regions and taking $1$ (a highly significant edge point) at sharp step edges. Generally, points at the same edge have a similar edge significance. As the steepness of a ramp edge reduces, the PAS values of the edge points also reduce. Thus, we can distinguish different ramp edges based on the PAS values of edge points. To overcome the deficiencies of the abovementioned algorithms and achieve a better despeckling performance both quantitatively and visually, we propose a Phase asymmetry Fractional anisotropic Diffusion and Total Variation (PFDTV) method to preserve various edges and remove speckle noises in ultrasound images.
	
The contributions of our method are four-fold. Firstly, to the best of our knowledge, this is the first fractional TV framework that leverages the PAS metric in designing weighting coefficients to maintain a balance between FAD and FTV filters in achieving not only the best despeckling performance with ramp edge preservation but also in reducing the staircase effect produced by integral-order filters. Secondly, a new diffusion coefficient of the FAD filter is proposed to properly preserve low-contrast edges by exploiting the edge identification of the PAS metric instead of intensity-based gradient information in traditional AD filters. Thirdly, our framework adjusts the fractional order adaptively based on the PAS metrics to enhance various edges: on the one hand, different from the fixed fractional-order diffusion filters \cite{46}\cite{47} that neglect the differences between smooth regions and the various edges in the spatially transitional border regions, our adaptive strategy adopts high-fractional order for the various edges and uses low-fractional order for smooth regions; on the other hand, our framework assigns an adaptive fractional order to each edge point according to the edge significance of the PAS metric for each edge point to obtain a better edge enhancement. Finaly, the proposed despeckling method is applied to remove complex cancer-dependent noises for ease of accurate breast cancer delineation in real ultrasound image segmentation. 

The remainder of this article is organized as follows. The theoretical backgrounds about phase asymmetry, fractional-order differential and fractional-order AD and TV filters are introduced in Section \ref{sec2}. The details of the PFDTV method are introduced in Section \ref{sec3}. The experimental results are reported in Section \ref{sec4}. Some research issues and future directions with a brief conclusion are summarized in Section \ref{sec5}. 

\section{THEORETICAL BACKGROUNDS} \label{sec2}
\subsection{Phase Asymmetry}\label{sec2.1}
For the characteristics of ultrasound images, solely employing gradient information to identify edges cannot achieve satisfactory feature preservation. This work detects edges by adopting the phase-based PAS measure, which can efficiently separate edges from smooth regions. According to a human perception study \cite{44}, at the points of perceivable step edges, the absolute values of even symmetric filter responses are small while the absolute values of odd symmetric filter responses are large. In other words, the difference between the odd and the even filter responses is large. According to this finding, PAS \cite{45} provides a good measure of the asymmetric image feature for detecting step and ramp-like edges.

To calculate the PAS metric of a 2D signal $f$, we first need to extract its local phase and local amplitude with a 2D signal decomposition method. In the past decades, there has been an increasing interest in decomposing multidimensional signals using spatially varying sinusoidal waves. As the understanding of the theory advanced, amplitude- and frequency-modulation (AM–FM) decompositions \cite{48}\cite{49}\cite{50} have been applied in a large range of problems, for example, ultrasound image texture analysis \cite{51}, ultrasound image segmentation \cite{40} and medical imaging \cite{52}. The monogenic signal \cite{53} was proposed to decompose the 2D signal $f$ into the local phase and local amplitude based on Riesz filters. The monogenic signal $f_{M}$ is defined as: $\mathop f\nolimits_M  = (f,\mathop f\nolimits_R ) = (f,\mathop r\nolimits_1  * f,\mathop r\nolimits_2  * f)$, where $f_{R}$ is the Riesz transform of $f$, $\mathop r\nolimits_1(x_{1},x_{2}) $ and $\mathop r\nolimits_2(x_{1},x_{2}) $ are the spatial representation of Riesz filters, shown as follows:
\begin{equation}
\begin{array}{*{20}{l}}
{{r_1}({\rm{ }}{x_1},{\rm{ }}{x_2}) = \frac{{ - {\rm{ }}{x_1}}}{{2\pi {{\left( {{\rm{ }}x_1^2 + {\rm{ }}x_2^2} \right)}^{3/2}}}}}\\
{{r_2}({\rm{ }}{x_1},{\rm{ }}{x_2}) = \frac{{ - {\rm{ }}{x_2}}}{{2\pi {{\left( {{\rm{ }}x_1^2 + {\rm{ }}x_2^2} \right)}^{3/2}}}}}
\end{array}
\end{equation}

Since natural images generally contain a wide range of frequencies, the monogenic signal $f_{M}$ needs to be combined with a set of bandpass quadrature filters $b$. The monogenic signal $f_{M}$ becomes $\text{ }{{f}_{M}}=\left( b*f,b*\text{ }{{r}_{1}}*f,b*\text{ }{{r}_{2}}*f \right)=({{e}_{s}},{{o}_{s}})$,  where ${e}_{s}$ and ${o}_{s}$ denote the scalar-valued even and vector-valued odd filter responses respectively.

Several families of bandpass filters $b$ have been proposed to calculate the ${e}_{s}$ and ${o}_{s}$; we adopt a Cauchy kernel as the bandpass filter, since the Cauchy kernel has an analytical expression in the spatial and the Fourier domain \cite{54}. In the frequency domain, the 2D isotropic Cauchy kernel is defined by the following:
\begin{equation}
C(w) = \mathop n\nolimits_c \mathop {\left| w \right|}\nolimits^a \exp \left( { - s\left| w \right|} \right),a \ge 1
\end{equation}
where $w = (\mathop w\nolimits_1 ,\mathop w\nolimits_2 )$ is the angular frequency, $s$ is the scaling parameter,  $\mathop n\nolimits_c  = \mathop {\left( {\frac{{\pi \mathop 4\nolimits^{a + 1} \mathop s\nolimits^{2a + 1} }}{{\Gamma (2a + 1)}}} \right)}\nolimits^{\frac{1}{2}} $, $\Gamma \left(  \cdot  \right)$ is the gamma function, and $a$ is the bandwidth. We set $a=1.58$, as suggested in \cite{39}.

To identify different edges accurately, Kovesi \cite{44} suggested to use the PAS measure over a number of scales. Therefore, we define the multiple-scale PAS as follows:
\begin{equation}
PA=\sum\limits_{s}{\frac{\left\lfloor \left| {{o}_{s}} \right|-\left| {{e}_{s}} \right|-\text{ }{{T}_{s}} \right\rfloor }{\sqrt{e_{s}^{2}+o_{s}^{2}}+\varepsilon }}
\end{equation}
where $PA$ is the PAS metric, $\varepsilon$ is a small positive constant to avoid division by zero, $T_{s}$ is the scale-specific noise threshold, $\left\lfloor  \cdot  \right\rfloor $ represents zeroing of negative values, and $s$ is the scaling parameter of Cauchy kernels. Specifically,  $s$ plays an important role in obtaining an accurate edge map, since increasing $s$ will regularize the continuity (or connect the breakpoints) in the boundaries but slightly lose details somewhat in edge detection. 

\begin{figure}	%插图
	\centering
	\includegraphics[width=\linewidth]{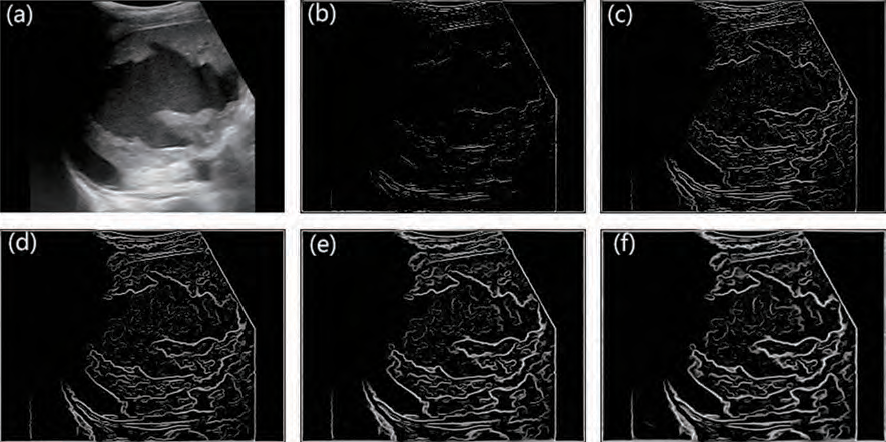}	%调节插图大小
	\caption{Example of PAS measure at different scales. (a) The ultrasound image of the spleen;  the PAS map of: (b) $s=5$,  (c) $s=10$, (d) $s=15$, (e) $s=20$, (f) $s=25$. }	%插图名称
	\label{fig1}	%各种label用来引用
\end{figure}

Fig. \ref{fig1} shows an example of the PAS measure at different scales. We can find that the discontinuities in some boundaries in the PAS maps at $s=5$ and $s=10$ will reduce the accuracy of locating edges. The boundaries in the PAS maps at $s=20$ and $s=25$ have good continuity, but some details are lost. The PAS map at $s=15$ maintains a balance between the boundary continuity and detail preservation. Thus, we choose $s=15$ to detect edges in real ultrasound images.   

PAS provides an absolute measure of the edge significance of points. The PAS metric varies from $0$ to $1$, taking $0$ (indicating non-edge significance) in ideal smooth regions and taking $1$ (indicating high edge significance) at sharp step edges. In general, points at the same edge have a similar edge significance. As the steepness of a step edge reduces, the PAS values of the edge points also reduce. Due to PAS being invariant to brightness or contrast, low-contrast edges can be detected efficiently. For the real ramp edges in ultrasound images, the PAS values of these edge points are less than $1$.

\subsection{Fractional-order differential}\label{sec2.2}
\begin{figure}	
	\centering
	\includegraphics[width=9cm]{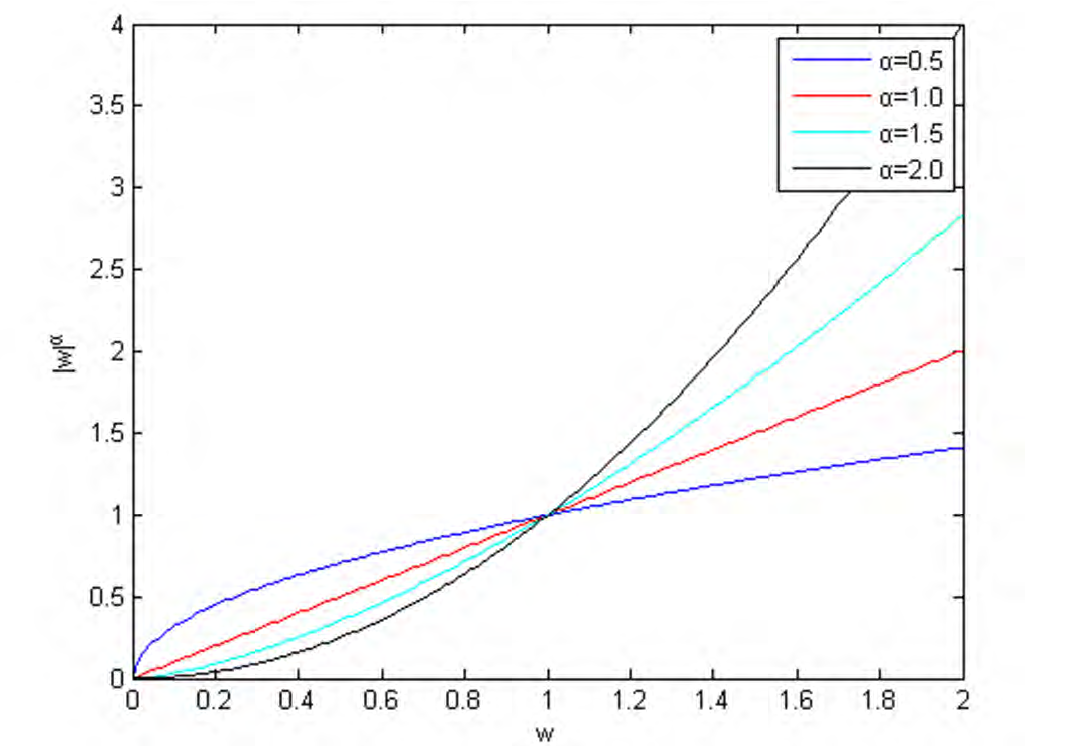}	
	\caption{The curves of the amplitude-frequency characteristic of the fractional differential with different orders.}	
	\label{fig2}	
\end{figure}

The fractional-order differential performs better in enhancing edges than integer-order differential during image processing \cite{47}. For a square differentiable signal $f(x) \in \mathop L\nolimits^2 (R) $, its fractional-order differential is given as follows:
\begin{equation}
\mathop D\nolimits^\alpha  f(x) = \frac{{\mathop d\nolimits^\alpha  f(x)}}{{d\mathop x\nolimits^\alpha  }}
\end{equation}
where $\alpha$ is a positive real number. The Fourier transform of $\mathop D\nolimits^\alpha  f(x)$ is as follows:
\begin{equation}
\begin{split}
\mathop D\nolimits^\alpha  f(x)&\mathop  \Leftrightarrow \limits^{FT} (\mathop {\hat D}\nolimits^\alpha  f)(w)= \mathop {(iw)}\nolimits^\alpha  \hat f(w) \\
&= \mathop {\left| w \right|}\nolimits^\alpha  \exp \left[ {i\mathop \theta \nolimits^\alpha  (w)} \right]\hat f(w)\\
&= \mathop {\left| w \right|}\nolimits^\alpha  \exp \left[ {\frac{{\alpha \pi i}}{2}{\mathop{\rm sgn}} (w)} \right]\hat f(w)
\end{split}
\end{equation}
where $w$ is the angular frequency, ${\mathop{\rm sgn}} ( \cdot )$ denotes the numeric symbol of the integer part,  and $\mathop {(iw)}\nolimits^\alpha   = \mathop {\left| w \right|}\nolimits^\alpha  \exp \left[ {\frac{{\alpha \pi i}}{2}{\mathop{\rm sgn}} (w)} \right]$ is the filter function of the fractional differential filter. According to the filter function, we can draw the curves of the amplitude-frequency characteristic of the fractional differential with different $\alpha$, as depicted in Fig. \ref{fig2}.
From Fig. \ref{fig2}, it is obviously seen that in the low-frequency field with $ 0<w<1$, the fractional differential acts as an attenuation function. Nevertheless, in the section with  $w> 1$, the fractional differential enlarges the amplitude values, and the enhanced amplitude will be stronger as the fractional order $\alpha$ increases. Taking into account the amplitude enhancement in the high-frequency field, we effectively apply the fractional-order differential into edge enhancement in image denoising.

Diffusion filters tend to reduce the edge contrast during smoothing. Although traditional fractional-order diffusion filters usually adopt one fixed fractional order to process the image, this strategy neglects the differences between smooth regions and the various edges in spatially transitional border regions \cite{47}. Edges will be weakened if a low-fractional order is used, while smooth regions will be ignored if a high-fractional order is adopted. This drawback will inevitably cause some details to be damaged after noise removal \cite{46}. Therefore, a more reasonable choice is to assign the fractional-order $\alpha$ adaptively based on the PAS metric as in Sec. \ref{sec3.2} for enhancing various edges in ultrasound images.  

Currently, there are three commonly used definitions of fractional calculus: the Capotu definition, the  Gr\"unwald–Letnikov (G-L) definition and the Riemann–Liouville (R-L) \cite{55}\cite{56}. Since the G-L definition expresses a function using the weighted sum around the function, the G-L definition is suitable for signal processing. According to \cite{57}, the $\alpha$-order differential of signal $ f(x) $ is defined by the G-L as follows:
\begin{equation}
D^{\alpha} f(x) \buildrel \Delta \over = \mathop {\lim }\limits_{h \to 0} \frac{1}{{{h^\alpha }}}\sum\limits_{l = 0}^{[\frac{{d - c}}{h}]} {{{( - 1)}^l}} \left(  \!\!\begin{array}{l}
\alpha \\
{l}
\end{array}  \!\!\right)f(x - lh)
\end{equation}
where $ \alpha $ is the fractional order, $ [c,d] $ is the duration of $ f(x) $, the integer part of $ \frac{d-c}{h} $ is $ [\frac{d-c}{h}] $, and the formula $  \left( \!\! \begin{array}{l}
\alpha \\
{l}
\end{array} \!\!\right) $ is the binomial coefficient defined as follows:
\begin{equation}
\left( \!\! \begin{array}{l}
\alpha \\
l
\end{array}  \!\!\right) = \frac{{\Gamma (\alpha  + 1)}}{{\Gamma (l + 1)\Gamma (\alpha  - l + 1)}}
\end{equation}
where $ \Gamma (n) = (n - 1)! $ is the gamma function.

\subsection{Fractional-order AD filter and fractional-order TV filter }
The following partial differential equation defines the original AD \cite{22} filter:
\begin{equation}
\frac{{\partial u}}{{\partial t}} = div\left[ {c\left( {\left| {\nabla u} \right|} \right) \cdot \nabla u} \right],
\end{equation}
where $ div $ is the divergence operator,  $ \left| {\nabla {u}} \right| $ is the absolute value of $ \nabla u $, and $c(\cdot)$ is the diffusion coefficient related to the magnitude of local image gradient $ \nabla u $. A possible diffusion coefficient function \cite{22} is defined as follows:
\begin{equation}
{c}\left( {\left| {\nabla u} \right|} \right) = 1/\left[ {1 + \mathop {\left| {\nabla u} \right|}\nolimits^2 /\mathop k\nolimits^2 } \right]
\end{equation}
where $k$ is the gradient threshold. To preserve edges, this diffusion coefficient will reduce the diffusivity at edges that have a large magnitude of the local intensity-based gradient.

The TV method for image processing is to search numerical approximation to the lowest cost function of image, the variational principle is the foundation of this method \cite{58}. Variational model for speckle noise removal \cite{59} normally consists of regularized term and data fidelity term. The classical TV filter \cite{58} is proposed as follows:
\begin{equation}
E(u) = \int_\Omega  {\left( {\left| {\nabla u} \right| + \frac{\lambda }{2}\mathop {\left| {u - \mathop u\nolimits_0 } \right|}\nolimits^2 } \right)}
\end{equation}
where $ \left| {\nabla {u}} \right|$ denotes the total variation, $\mathop {\left| {u - \mathop u\nolimits_0 } \right|}\nolimits^2 $ is the fidelity term, $ \lambda $ is the regularization parameter and $ u_{0} $ is the noisy image. 

To reduce the staircase effect caused by classical AD and TV filters, the variational FAD \cite{28} and FTV \cite{60} filters were developed. The variational FAD filter \cite{28} is shown as follows:
\begin{equation}
E\left( u \right) = \int_\Omega  {f\left( {\left| {\mathop \nabla \nolimits^\alpha  u} \right|} \right)} d\Omega
\end{equation}
where $\alpha$ is the fractional order, $ \mathop \nabla \nolimits^\alpha  u = \left( {\mathop \nabla \nolimits_x^\alpha  u,\mathop \nabla \nolimits_y^\alpha u } \right) $, $ \left| {\mathop \nabla \nolimits^\alpha  u} \right| = \sqrt {\mathop {\left( {\mathop\nabla \nolimits_x^\alpha  u} \right)}\nolimits^2  + \mathop {\left( {\mathop \nabla \nolimits_y^\alpha u } \right)}\nolimits^2 } $,  and $ f\left( {\left| {\mathop \nabla \nolimits^\alpha  u} \right|} \right) \ge 0 $ is an increasing function associated with the diffusion coefficient in the AD \cite{22} filter, shown as follows:
\begin{equation}
c\left( t \right) = \frac{{f'\left( {\sqrt t } \right)}}{{\sqrt t }}
\end{equation}
Zhang and Wei \cite{60} proposed the following FTV filter:
\begin{equation}
E\left( u \right) = \int_\Omega  {\left( {\left| {\mathop \nabla \nolimits^\alpha  u} \right| + \frac{\lambda }{2}\mathop {\left| {u - \mathop u\nolimits_0 } \right|}\nolimits^2 } \right)} dxdy
\end{equation}
where $\alpha$ is the fractional order, $\left| {\mathop \nabla \nolimits^\alpha  u} \right|$ denotes the total variation, $\mathop {\left| {u - \mathop u\nolimits_0 } \right|}\nolimits^2 $ is the fidelity term, $ \lambda $ is the regularization parameter and $ u_{0} $ is the noisy image.

\section{OVERALL OF PFDTV} \label{sec3}
\subsection{The proposed model}\label{sec3.2}
The PFDTV method proposes a new TV framework with phase asymmetry to guide adaptive fractional-order total variation and diffusion filter for feature-preserving ultrasound despeckling. Specifically, the proposed TV cost function combining equation (11) and (13) balances the FAD filter and FTV filter for achieving the best performance of preserving ramp edges, and the TV cost function is defined as follows:
\begin{equation}
E\left( u \right) = \int_\Omega  {\left[ {\varphi f\left( {\left| {\mathop \nabla \nolimits^{\alpha } u} \right|} \right) + \gamma \left| {\mathop \nabla \nolimits^{\alpha } u} \right| + \frac{\lambda }{2}\mathop {\left| {u - \mathop u\nolimits_0 } \right|}\nolimits^2 } \right]} dxdy
\end{equation}
where  $\alpha$ is the adaptive fractional order, $\varphi$ and $\gamma$ are the weighted coefficients that control the relative importance of the FAD and FTV filters, and $\lambda$ is the regularization parameter. The speckle filtering result is closely related to $\lambda$: the greater the value of $\lambda$ is, the more noisy the filtered image achieves; the smaller the value of $\lambda$ is, the more blurry the edge feature becomes. Two types of regularization parameter selection methods have been introduced in terms of locally-adaptive and global performance measures. In this paper, the parameter $\lambda$ is chosen so that the best global PSNR of despeckled image is obtained \cite{60}. We empirically set $\lambda$ as $0.01$ after synthetic image despeckling experiments. Regarding the weighted coefficients, we design them based on the PAS metric shown as follows:
\begin{equation}
\left\{ \begin{array}{l}
\varphi = \mathop {(PA - 1)}\nolimits^2 \\
\gamma  = PA(2 - PA)
\end{array} \right.
\end{equation}
where $ PA $ is the PA metric that is updated in each iteration for accurately obtaining the edge significance of each point.

Based on the above strategy, when $PA$ is close to $0$, we emphasize the role of the FAD filter in smoothing regions. When $PA$ is close to $1$, we highlight the role of the FTV filter in the boundary regions for edge preservation. Because the $PA$ values of edge points for real ramp edges are less than $1$, the FAD filter also plays a key role in processing these weak edges in the spatially transitional border regions. 

However, the FAD filter solely integrates intensity-based gradient into the diffusion coefficient, causing some low-contrast edges to be removed after noise removal. To overcome this drawback, we integrate the PAS metric into the diffusion coefficient. The PAS measure can efficiently identify low-contrast edges due to its invariance to brightness or contrast. Furthermore, the PAS metric value is only related to the edge significance of each point. We alter the function $f\left( {\left| {\mathop \nabla \nolimits^\alpha  u} \right|} \right)$ by modifying its diffusion coefficient $c( \cdot )$ according to (12). The modified diffusion coefficient is shown as follows:
\begin{equation}
c\left( \left| {{\nabla }^{\alpha }}u \right|,PA \right)=1/\left[ 1+\frac{\left| {{\nabla }^{\alpha }}u \right|\cdot \left( 1+254\cdot PA \right)}{\text{ }k_{1}^{2}} \right]
\end{equation}
where $\mathop k\nolimits_1  = \mathop k\nolimits_0 \mathop e\nolimits^{ - 0.05(n_{iter} - 1)} $ is the modified version of $k$ in (9). Here, $n_{iter}$ is the number of iterations, and $k_{0}$ is a positive constant that is related to the noise level.

Given that diffusion filters reduce the edge contrast during smoothing, it is essential to design a proper strategy to enhance various edges in an ultrasound image. According to the discussion in Sec. \ref{sec2.2}, we can set the fractional-order $\alpha$ as a monotone increasing function of the PAS value of the ultrasound images to adaptively enhance the various edges in the ultrasound images. Specifically, we adopt a logarithmic function to describe the adaptive adjustment of the order of fractional derivative by modifying the adaptive fractional-order strategy in \cite{46} as follows:
\begin{equation}
\alpha = 1 + \mathop {\log }\nolimits_2 \left( {1 + PA^2} \right)
\end{equation}
where $ PA $ is the PAS metric. This functional setting ensures that $\alpha \in \left( 1,2 \right)$. The adaptive strategy of the PFDTV method assigns low-fractional order to preserve smooth regions and uses high-fractional order to enhance the various edges in spatially transitional border regions. 

In fact, the PAS metric shows the edge significance of each point. As the PAS value increases, the edge significance of the point also increases and the point is more likely to be an edge point. According to (17), a larger PAS metric yields a larger $\alpha$ that can produce better edge enhancement. The PFDTV method will adopt relatively high-fractional order to enhance the most significant edge points compared with the least significant edge points, so that we can properly preserve various edges and obtain a better image enhancement.

\subsection{Numerical Solver}
We leverage the Euler-Lagrange equation \cite{61} to solve the cost function (14). Assuming the solution $u$ of this cost function $E(u)$ is known, then this solution must make $E(u)$ minimum. In other words, adding any slight perturbation to $u$ will make the cost function larger. When the perturbation goes to 0, the derivative of the cost function with respect to the perturbation is 0. The perturbation is represented as a very small continuous function $\eta  \in \mathop C\nolimits^\infty  \left( \Omega  \right)$ multiplied by a perturbation factor $e$. Define the following:
\begin{equation}
\begin{array}{l}
\Phi \left( {e} \right): = E\left( {u + e\eta } \right)\\
= \int_\Omega  {\left[ {\varphi f\left( {\left| {\mathop \nabla \nolimits^\alpha  \left( {u + e\eta } \right)} \right|} \right) + \gamma \left| {\mathop \nabla \nolimits^\alpha  \left( {u + e\eta } \right)} \right|} \right]} dxdy\\
+ \int_\Omega  {\left( {\frac{\lambda }{2}\mathop {\left| {u + e\eta  - \mathop u\nolimits_0 } \right|}\nolimits^2 } \right)} dxdy
\end{array}
\end{equation}
We first take the derivative of $\Phi(e)$ and obtain the following:
\begin{equation}
\begin{array}{l}
\Phi '\left( e \right)  =\frac{d}{{de}}\Phi \left( e \right) =\\
\varphi \int_\Omega  {\left( {f'\left( {\left| {\mathop \nabla \nolimits^\alpha  \left( {u + e\eta } \right)} \right|} \right)  \frac{{\mathop \nabla \nolimits_x^\alpha  \left( {u + e\eta } \right)\mathop \nabla \nolimits_x^\alpha  \eta  + \mathop \nabla \nolimits_y^\alpha  \left( {u + e\eta } \right)\mathop \nabla \nolimits_y^\alpha  \eta }}{{\sqrt {\mathop {\left( {\mathop \nabla \nolimits_x^\alpha  \left( {u + e\eta } \right)} \right)}\nolimits^2  + \mathop {\left( {\mathop \nabla \nolimits_y^\alpha  \left( {u + e\eta } \right)} \right)}\nolimits^2 } }}} \right)} dxdy\\
+ \gamma \int_\Omega  {\left( {\frac{{\mathop \nabla \nolimits_x^\alpha  \left( {u + e\eta } \right)\mathop \nabla \nolimits_x^\alpha  \eta  + \mathop \nabla \nolimits_y^\alpha  \left( {u + e\eta } \right)\mathop \nabla \nolimits_y^\alpha  \eta }}{{\sqrt {\mathop {\left( {\mathop \nabla \nolimits_x^\alpha  \left( {u + e\eta } \right)} \right)}\nolimits^2  + \mathop {\left( {\mathop \nabla \nolimits_y^\alpha  \left( {u + e\eta } \right)} \right)}\nolimits^2 } }}} \right)} dxdy\\
+ \lambda \int_\Omega  {\left( {u + e\eta  - \mathop u\nolimits_0 } \right)}\eta dxdy,
\end{array}
\end{equation}
Let $e=0$, and we have the following:
\begin{equation}
\begin{array}{l}
\Phi '\left( 0 \right) =\\
 \varphi \int_\Omega  {\left( {c\left( {\mathop {\left| {\mathop \nabla \nolimits^\alpha  u} \right|}\nolimits^2, PA^2 } \right) \left( {\mathop \nabla \nolimits_x^\alpha  u\mathop\nabla \nolimits_x^\alpha  \eta  + \mathop \nabla \nolimits_y^\alpha  u\mathop \nabla \nolimits_y^\alpha  \eta } \right)} \right)} dxdy\\
+ \gamma \int_\Omega  {\frac{{\mathop\nabla \nolimits_x^\alpha  u\mathop\nabla \nolimits_x^\alpha  \eta  + \mathop \nabla \nolimits_y^\alpha  u\mathop \nabla \nolimits_y^\alpha  \eta }}{{\left| {\mathop \nabla \nolimits^\alpha  u} \right|}}} dxdy\\
+ \lambda \int_\Omega  {(u - \mathop u\nolimits_0 )}\eta dxdy
\end{array}
\end{equation}
where $ \left| {\mathop \nabla \nolimits^\alpha  {u}} \right| = \sqrt {\mathop {\left( {\mathop \nabla \nolimits_x^\alpha u } \right)}\nolimits^2  + \mathop {\left( {\mathop \nabla \nolimits_y^\alpha u } \right)}\nolimits^2 }  $. According to the previous analysis for finding the solution $u$, we can obtain $\Phi '\left( 0 \right) =0$.
To simplify (20), we use the definition of the adjoint operator to simplify the following term:
\begin{equation}
\mathop \nabla \nolimits_x^\alpha  u\mathop \nabla \nolimits_x^\alpha  \eta  + \mathop \nabla \nolimits_y^\alpha  u\mathop \nabla \nolimits_y^\alpha  \eta  = \left( {\mathop {\left( {\mathop \nabla \nolimits_x^\alpha  } \right)}\nolimits^ *  \mathop \nabla \nolimits_x^\alpha  u + \mathop {\left( {\mathop \nabla \nolimits_y^\alpha  } \right)}\nolimits^ *  \mathop \nabla \nolimits_y^\alpha  u} \right)\eta
\end{equation}
where $ {\mathop {\left( {\mathop \nabla \nolimits_x^\alpha  } \right)}\nolimits^ *  } $ and $ {\mathop {\left( {\mathop \nabla \nolimits_y^\alpha  } \right)}\nolimits^ *  } $ are the adjoint operators of $\mathop \nabla \nolimits_x^\alpha $ and $ \mathop \nabla \nolimits_y^\alpha $ respectively \cite{62}. Based on the above analysis, we obtain the simplified form of (20) as follows:
\begin{equation}
\begin{array}{l}
\Phi '\left( 0 \right){\rm{ = }}\\
\varphi \int_\Omega  {c\left( {\mathop {\left| {\mathop \nabla \nolimits^\alpha  u} \right|}\nolimits^2,PA^2 } \right)\left( {\mathop {\left( {\mathop \nabla \nolimits_x^\alpha  } \right)}\nolimits^ *  \mathop \nabla \nolimits_x^\alpha  u + \mathop {\left( {\mathop \nabla \nolimits_y^\alpha  } \right)}\nolimits^ *  \mathop \nabla \nolimits_y^\alpha  u} \right)\eta } {\rm{dx}}dy\\
+ \gamma \int_\Omega  {\frac{{\mathop {\left( {\mathop \nabla \nolimits_x^\alpha  } \right)}\nolimits^ *  \mathop \nabla \nolimits_x^\alpha  u + \mathop {\left( {\mathop \nabla \nolimits_y^\alpha  } \right)}\nolimits^ *  \mathop \nabla \nolimits_y^\alpha  u}}{{\left| {\mathop \nabla \nolimits^\alpha  u} \right|}}\eta } {\rm{dx}}dy\\
+ \lambda \int_\Omega  {\left( {u - \mathop u\nolimits_0 } \right)\eta } {\rm{dx}}dy
\end{array}
\end{equation}
For all $\eta  \in \mathop C\nolimits^\infty  \left( \Omega  \right)$, the Euler-Lagrange equation is:
\begin{equation}
\begin{array}{l}
\varphi c\left( {\mathop {\left| {\mathop \nabla \nolimits^\alpha  u} \right|}\nolimits^2, PA^2} \right)\left( {\mathop {\left( {\mathop \nabla \nolimits_x^\alpha  } \right)}\nolimits^ *  \mathop \nabla \nolimits_x^\alpha  u + \mathop {\left( {\mathop \nabla \nolimits_y^\alpha  } \right)}\nolimits^ *  \mathop \nabla \nolimits_y^\alpha  u} \right) \\
+\gamma \frac{{\mathop {\left( {\mathop\nabla \nolimits_x^\alpha  } \right)}\nolimits^ *  \mathop \nabla \nolimits_x^\alpha  u + \mathop {\left( {\mathop \nabla \nolimits_y^\alpha  } \right)}\nolimits^ *  \mathop \nabla \nolimits_y^\alpha  u}}{{\left| {\mathop \nabla \nolimits^\alpha  u} \right|}} + \lambda \left( {u - \mathop u\nolimits_0 } \right) = 0
\end{array}
\end{equation}
where $u$ is the solution that minimizes the cost function.

Let $\nabla E $ denote the first derivative of the cost function $E(u)$; a necessary condition for $u$ to be the extreme point of $E(u)$ is that $\nabla E=0$. Thus, $\nabla E$ holds that:
\begin{equation}
\begin{array}{l}
\nabla E = \varphi c(\mathop {\left| {\mathop \nabla \nolimits^\alpha  u} \right|}\nolimits^2,PA^2 )(\mathop {(\mathop \nabla \nolimits_x^\alpha  )}\nolimits^ *  \mathop \nabla \nolimits_x^\alpha  u + \mathop {(\mathop \nabla \nolimits_y^\alpha  )}\nolimits^ *  \mathop \nabla \nolimits_y^\alpha  u) \\
+ \gamma \frac{{\mathop {(\mathop \nabla \nolimits_x^\alpha  )}\nolimits^ *  \mathop \nabla \nolimits_x^\alpha  u + \mathop {(\mathop \nabla \nolimits_y^\alpha  )}\nolimits^ *  \mathop \nabla \nolimits_y^\alpha  u}}{{\left| {\mathop \nabla \nolimits^\alpha  u} \right|}} + \lambda (u - \mathop u\nolimits_0 )
\end{array}
\end{equation}

The desired $u$ is computed via gradient descent method as in \cite{63}. Specifically, we introduce an artificial time parameter $ \Delta t $ and take a small step in the direction of $-\nabla E$, $i.e.$, $u^{n+1}=u^{n}+\Delta t(-\nabla E)$.
Finally, we will obtain the desired image $u$ that minimizes the cost function $E(u)$.

\subsection{Numerical algorithm}
To compute (24) numerically, we use the G-L fractional differential to facilitate the numerical implementation. We assume that the size of a given image $u$ is $ X \times Y $, where $ X $ and $ Y $ are the numbers of pixels in the vertical and horizontal direction respectively. Then, we can obtain the discretized schemes  of $\mathop \nabla  \nolimits_{\rm{x}}^\alpha$, $\mathop \nabla  \nolimits_y^\alpha  $ , $\mathop {\left( {\mathop \nabla  \nolimits_x^\alpha  } \right)}\nolimits^ *  $ and $\mathop {\left( {\mathop \nabla  \nolimits_y^\alpha  } \right)}\nolimits^ *  $, shown as follows:
\begin{equation}
\left\{ \begin{array}{l}
\mathop\nabla  \nolimits_x^\alpha  \mathop u\nolimits_{i,j}  = \sum\limits_{l = 0}^j {\mathop {\left( { - 1} \right)}\nolimits^l \left( \!\!\begin{array}{l}
	\alpha \\
	l
	\end{array} \!\!\right)} \mathop u\nolimits_{i,j - l} \\
\mathop \nabla  \nolimits_y^\alpha  \mathop u\nolimits_{i,j}  = \sum\limits_{l = 0}^i {\mathop {\left( { - 1} \right)}\nolimits^l \left(\!\! \begin{array}{l}
	\alpha \\
	l
	\end{array} \!\!\right)\mathop u\nolimits_{i - l,j} }
\end{array} \right.
\end{equation}
\begin{equation}
\left\{ \begin{array}{l}
\mathop {\left( {\mathop\nabla  \nolimits_x^\alpha  } \right)}\nolimits^ *  \mathop u\nolimits_{i,j}  = \sum\limits_{l = 0}^{Y - 1 - j} {\mathop {\left( { - 1} \right)}\nolimits^l \left(\!\! \begin{array}{l}
	\alpha \\
	l
	\end{array}\!\! \right)} \mathop u\nolimits_{i,j + l} \\
\mathop {\left( {\mathop \nabla \nolimits_y^\alpha  } \right)}\nolimits^ *  \mathop u\nolimits_{i,j}  = \sum\limits_{l = 0}^{X - 1 - i} {\mathop {\left( { - 1} \right)}\nolimits^l \left(\!\! \begin{array}{l}
	\alpha \\
	l
	\end{array} \!\!\right)\mathop u\nolimits_{i + l,j} }
\end{array} \right.
\end{equation}
where $i=0,1,...,X-1, j=0,1,...,Y-1$, and the formula $\left( \!\!\begin{array}{l}
\alpha \\
{l}
\end{array} \!\!\right)$ is the binomial coefficient defined as
$\left(\!\! \begin{array}{l}
\alpha \\
{l}
\end{array}\!\! \right) = \frac{{\Gamma \left( {\alpha  + 1} \right)}}{{\Gamma (l + 1)\Gamma (\alpha  - l + 1)}}$, where $\Gamma $ is the gamma function. Let the following:
\begin{equation}
\!\!\left\{ {\begin{array}{*{20}{l}}
\!\!\!\!\!	{FA{D_x}{\rm{ }}{u_{i,j}}\!\! =\!\!\! \sum\limits_{l = 0}^{Y - 1 - j}\!\!\! {{{({\rm{ }} - {\rm{ }}1)}^l}\left(\!\! {\begin{array}{*{20}{l}}
				\alpha \\
				l
				\end{array}}\!\! \right)} \frac{{{\rm{ }}k_1^2\nabla _x^\alpha {\rm{ }}{u_{i,j + l}}}}{{{\rm{ }}k_1^2 + {{\left| {{\nabla ^\alpha }{\rm{ }}{u_{i,j + l}}} \right|}^2}{{\left[ {1 + 254PA(\mathop u\nolimits_{i,j + l} )} \right]}^2}}}}\\
\!\!\!\!\!	{FA{D_y}{\rm{ }}{u_{i,j}}\!\! = \!\!\!\sum\limits_{l = 0}^{X - 1 - i}\!\!\! {{{({\rm{ }} - {\rm{ }}1)}^l}\left( \!\!{\begin{array}{*{20}{l}}
				\alpha \\
				l
				\end{array}}\!\! \right)} \frac{{{\rm{ }}k_1^2\nabla _y^\alpha {\rm{ }}{u_{i + l,j}}}}{{{\rm{ }}k_1^2 + {{\left| {{\nabla ^\alpha }{\rm{ }}{u_{i + l,j}}} \right|}^2}{{\left[ {1 + 254PA\left( {\mathop u\nolimits_{i + l,j} } \right)} \right]}^2}}}}
	\end{array}} \right.
\end{equation}
\begin{equation}
\!\!\left\{ \begin{array}{l}
\!\!\!\!\!\mathop {FTV}\nolimits_x \mathop u\nolimits_{i,j}\!\!  =\!\!\! \sum\limits_{l = 0}^{Y - 1 - j}\!\! {\mathop {\left( { - 1} \right)}\nolimits^l \left(\!\! \begin{array}{l}
	\alpha \\
	l
	\end{array}\!\! \right)} \frac{{\mathop \nabla \nolimits_x^\alpha  \mathop u\nolimits_{i,j + l} }}{{\sqrt {\mathop {\left( {\mathop \nabla \nolimits_x^\alpha  \mathop u\nolimits_{i,j + l} } \right)}\nolimits^2  + \mathop {\left( {\mathop \nabla \nolimits_y^\alpha  \mathop u\nolimits_{i,j + l} } \right)}\nolimits^2  + \varepsilon } }}\\
\!\!\!\!\!\mathop {FTV}\nolimits_y \mathop u\nolimits_{i,j} \!\! =\!\!\! \sum\limits_{l = 0}^{X - 1 - i}\!\! {\mathop {\left( { - 1} \right)}\nolimits^l \left(\!\! \begin{array}{l}
	\alpha \\
	l
	\end{array} \!\!\right)} \frac{{\mathop \nabla \nolimits_y^\alpha  \mathop u\nolimits_{i+ l,j } }}{{\sqrt {\mathop {\left( {\mathop \nabla \nolimits_x^\alpha  \mathop u\nolimits_{i+ l,j } } \right)}\nolimits^2  + \mathop {\left( {\mathop\nabla \nolimits_y^\alpha  \mathop u\nolimits_{i+ l,j } } \right)}\nolimits^2  + \varepsilon } }}
\end{array} \right.
\end{equation}
where $PA(u_{i,j})$ is the PAS value of pixel $u_{i,j}$, and $ \varepsilon $ is very small positive number. We summarize the optimization process in Algorithm \ref{alg1}.
\begin{algorithm}
	\renewcommand{\algorithmicrequire}{\textbf{Input:}}
	\renewcommand{\algorithmicensure}{\textbf{Output:}}
	\caption{PFDTV feature-preserving despeckle filter}
	\label{alg1}
	\begin{algorithmic}[1]
		\REQUIRE noisy ultrasound image $u_{0}$, the values of $s$, $k_{0}$, time step $\Delta t$ and iteration number $n_{iter}$
		\ENSURE the despeckled image $u$
		\STATE Initialize $ u^{(0)}=u_{0}, \varepsilon=0.0001, n=1$, 
		\FORALL{$n<n_{iter}$}
		\STATE Compute $\mathop {FAD}\nolimits_x \mathop u\nolimits_{i,j}^{(n)} ,\mathop {FAD}\nolimits_y \mathop u\nolimits_{i,j}^{(n)} ,\mathop {FTV}\nolimits_x \mathop u\nolimits_{i,j}^{(n)} $ and $\mathop {FTV}\nolimits_y \mathop u\nolimits_{i,j}^{(n)} $ using (27) and (28)
		\STATE Compute $u^{(n+1)}$ through the following procedure:\\ $\mathop u\nolimits_{i,j}^{(n + 1)}  = \mathop u\nolimits_{i,j}^{(n)}  - \Delta t[ \varphi (\mathop {FAD}\nolimits_x \mathop u\nolimits_{i,j}^{(n)}  + \mathop {FAD}\nolimits_y \mathop u\nolimits_{i,j}^{(n)} )
		+ \gamma (\mathop {FTV}\nolimits_x \mathop u\nolimits_{i,j}^{(n)}  + \mathop {FTV}\nolimits_y \mathop u\nolimits_{i,j}^{(n)} )  + \lambda (\mathop u\nolimits_{i,j}^{(n)}  - \mathop u\nolimits_{i,j}^{(0)} )]$
		\STATE Set $n=n+1$
		\ENDFOR
		\STATE Set the despeckled image $u=u^{(n)}$
		\STATE \textbf{return} $u$
	\end{algorithmic}  
\end{algorithm}

\section{EXPERIMENTAL RESULTS} \label{sec4}
Experiments with synthetic and clinical ultrasound images were carried out to show the performance of the proposed PFDTV method\footnote{The source code is available at https://github.com/Binjie-Qin/PFDTV}. Several well-known ultrasound despeckling filters were used for comparison, including Frost \cite{14}, SRAD \cite{23}, OBNLM \cite{18}, SBF \cite{17}, ADLG \cite{8}, and NLLRF \cite{20}. We directly requested the source code of the SBF filter from its authors. In regard to other filters, we obtained the source codes from the cited authors' websites listed in their works.
 
\subsection{Synthetic image experiment}
\begin{figure}	%插图
	\centering
	\includegraphics[width=9cm]{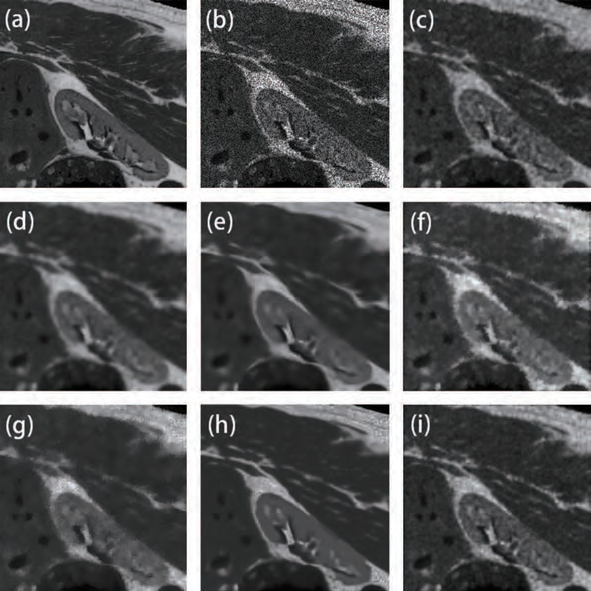}	
	\caption{Results of different filters for the synthetic image. (a) Original image, (b) original image corrupted with speckle noise with variance $\sigma^{2}=0.2$, despeckled result by (c) Frost ($W = 5 \times 5$) , (d) SRAD ($\Delta t = 0.3,\mathop n\nolimits_{iter}  = 60$), (e) OBNLM ($M = 5,\alpha  = 6,h = 3$) , (f) SBF ($W{\rm{ = }}3 \times 3,\mathop {\rm{n}}\nolimits_{{\rm{iter}}}  = 9$), (g) ADLG ($\Delta {\rm{t}} = 0.3,\mathop n\nolimits_{iter}  = 50$), (h) NLLRF ($\beta=20,H=10$), and (i) \textbf{PFDTV} ($\Delta t = 0.3,s=20,\mathop k\nolimits_0  = 100 , \mathop n\nolimits_{iter}  = 7$). }	%插图名称
	\label{fig3}	%各种label用来引用
\end{figure}

For the purpose of quantitative comparisons, we generated noise over the ground truth image by employing the synthetic speckle noise model that is widely used in the literature \cite{18}\cite{20}\cite{39}. The noise model is given by the following:
 \begin{equation}
 u(x_{i}) = v(x_{i}) + v(x_{i})\tau (x_{i}),\tau (x_{i}) \sim N(0,\mathop \sigma \nolimits^2 )
 \end{equation}
 where $v(x_{i})$ and $u(x_{i})$ are the pixel intensities of pixel $x_{i}$ in the
 noise-free image and the synthesized noisy image respectively, and $\tau(x_{i})$ is a zero-mean Gaussian noise with variance $\sigma^{2}$. We applied this noise model to Fig. \ref{fig3}(a), which consists of smooth regions and various local features. Three levels of noise were tested by setting $ \mathop \sigma \nolimits^2 {\rm{ = }}\left\{ {0.2;0.4;0.6} \right\} $. Fig. \ref{fig3}(b) depicts the synthetic image with noise variance $\mathop \sigma \nolimits^2 {\rm{ = }}0.2$.
\begin{table*}
	\centering
	\fontsize{9}{14}\selectfont
	\caption{Comparison of the PSNR, MSSIM and FSIM values among different filters.}
	\label{tab1}
	\begin{tabular}{lccccccccc}
		\hline
		\multirow{2}{*}{} &
		\multicolumn{3}{c}{PSNR} &
		\multicolumn{3}{c}{MSSIM} &
		\multicolumn{3}{c}{FSIM}\cr
		\cline{2-10} % \cline用于画横线 \cline{i-j}表示从第i列画到第j列
		& 0.2 & 0.4 & 0.6&0.2&0.4&0.6&0.2&0.4&0.6 \\
		\hline
		Frost&25.773&23.840&22.955&0.721&0.638&0.576&0.848&0.818&0.797 \cr
		SRAD&26.760&25.272&24.282&0.743&0.697&0.679&0.816&0.777&0.764\cr
		OBNLM &26.496&24.974&24.121&0.742&0.696&0.679&0.822&0.794&0.764\cr
		SBF&22.502&22.303&22.214&0.695&0.648&0.657&0.825&0.807&0.790\cr
		ADLG&24.585&22.945&22.360&0.676&0.625&0.599&0.786&0.758&0.745\cr
		NLLRF&27.409&25.536&24.282&0.763&0.711&0.699&0.825&0.798&0.768\cr
		\textbf{PFDTV}&{\bf 27.721}&{\bf 25.994}&{\bf 25.103}&{\bf 0.783}&{\bf 0.732}&{\bf 0.713}&{\bf 0.867}&{\bf 0.844}&{\bf 0.834}\cr\hline
	\end{tabular}
\end{table*}
To quantitatively evaluate the performance of each filter, the peak signal-to-noise ratio (PSNR) \cite{64}\cite{65}, mean structural similarity (MSSIM) \cite{66} and feature similarity index (FSIM) \cite{67} were adopted in this paper. Specifically, FSIM\footnote{\url{ http://sse.tongji.edu.cn/linzhang/IQA/FSIM/FSIM.htm}} is designed for measuring the ability of preserving features, and it takes values between 0 and 1, and 1 denotes the best performance of feature preservation.

To fairly compare the overall performance of each filter in noise reduction and feature preservation, each filter needs to achieve the best performance by setting its optimal parameters. The optimal parameters need to be selected based on a quantitative metric so that the PSNR metric is accepted as a gold standard as in \cite{2}\cite{68}. This is because PSNR regards the structural information and the nonstructural information as the same in terms of the contribution towards the performance, whereas SSIM and FSIM put more emphasis on the structural information \cite{2}. Therefore, we obtain the optimal parameters of each filter when achieving the highest value of the PSNR metric. PFDTV has the following parameters: $s$ is the scale of the Cauchy kernel, $\Delta t$ is the time step, $n_{iter}$ is the iteration number, and $k_{0}$ is for the noise level.

Fig. \ref{fig3} depicts the denoised images of different filters with their optimal parameters. The Frost filter has clear features, but it retains a significant level of noise. The SRAD remove noise better but produces smoother edges and removes some low-contrast features compared with Frost. Though OBNLM and NLLRF have good performance in high-contrast features, they cause some meaningful low-contrast features to become indiscernible after noise removal. Both SBF and ADLG produce fuzzy boundaries, and ADLG removes considerable details. Therefore, the PFDTV method achieves the best performances in noise removal and feature preservation.

Table \ref{tab1} compares the PSNR, MSSIM and FSIM values for different filters. The proposed PFDTV method achieves the highest PNSR, MSSIM and FSIM. The highest PSNR denotes that the despeckled image of our method produces a lower image distortion compared with other filters. The highest MSSIM represents that the despeckled image of our method is most similar to the original image. The highest FSIM represents that the PFDTV method outperforms other filters in feature preservation.

\subsection{Clinical image experiment}
Since real ultrasound images are all affected by speckle noise, there are no ground truth image and gold standard for quantitative performance evaluation on real ultrasound images. Therefore, we employed different types of clinical ultrasound images to visually verify the performance of the PFDTV method. The clinical ultrasound images were all downloaded from the public dataset.\footnote{\url{http://www.ultrasoundcases.info}}
\begin{figure}	%插图
	\centering
	\includegraphics[width=8cm]{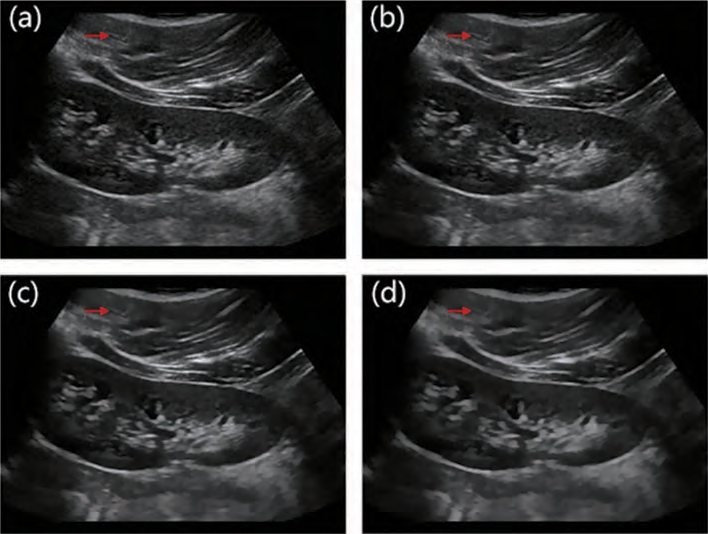}	
	\caption{Example of the despeckled image of different $k_{0}$. (a) The original ultrasound image; the despeckled result of (b) $k_{0}=5$, (c) $k_{0}=20$, (d) $k_{0}=100$.}	%插图名称
	\label{fig4}	%各种label用来引用
\end{figure}

The real ultrasound image experiment has no ground truth image and gold standard to find the optimal parameters in terms of quantifiable performance criteria. Thus, we adjust the parameters of the PFDTV method to obtain the best visual effect. The visual performance of our method is closely associated with three parameters: $s$, $k_{0}$ and $n_{iter}$. According to the analysis in Section \ref{sec2.1}, we set the scale $s$ of the PAS measure as $15$. $k_{0}$ is a positive constant that is related to the the noise level, and $n_{iter}$ is the number of iterations.

The parameter $k_{0}$ is an important parameter that is related to the noise level estimation. Noise level estimation is a prerequisite for various detail-preserving image denoising methods \cite{21} but has remained as an unsolved challenging problem in ultrasound image despeckling. According to (16), if $k_{0}$ is set to a large value, a large diffusion coefficient is obtained that will cause some details to be damaged after speckle reduction. However, if $k_{0}$ is set to a small value, a small diffusion coefficient is achieved that will increase computational costs in the despeckling process. To explore the impact of parameter $k_{0}$, we set $k_{0}$ with different values. Fig. \ref{fig4} depicts the despeckled results of different $k_{0}$. It is observed that there is a significant level of noise in the despeckled result of $k_{0}=5$. As indicated by the red arrow in Fig. \ref{fig4}(d), the meaningful detail is damaged after noise removal in the case of $k_{0}=100$. Our method achieves not only satisfactory feature preservation but also desirable noise reduction in the case of $k_{0}=20$. In real image experiments, we find that the best range of $k_{0}$ is from $20$ to $40$. Here, we set $k_{0}$ as $20$.
\begin{figure}	%插图
	\centering
	\includegraphics[width=8cm]{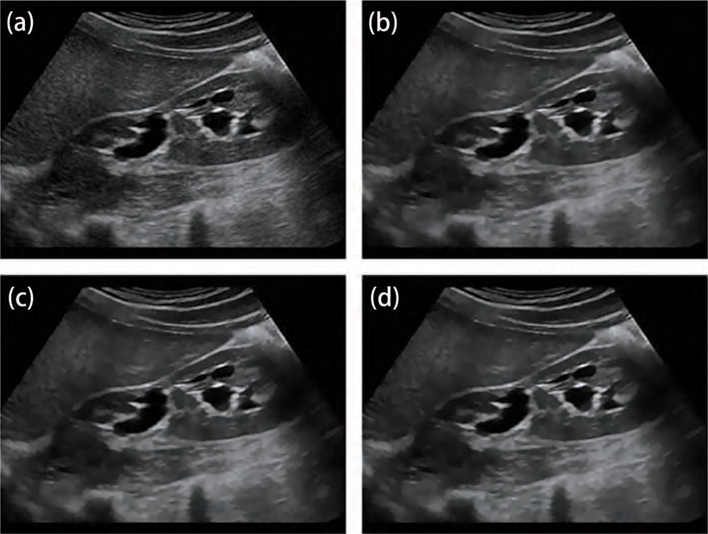}	
	\caption{Example of the despeckled images of different $n_{iter}$. (a) The original ultrasound image; the despeckled result of (b) $n_{iter}=8$, (c) $n_{iter}=25$, and (d) $n_{iter}=50$.}	%插图名称
	\label{fig5}	%各种label用来引用
\end{figure}
\begin{figure*}	%插图
	\centering
	\includegraphics[width=\linewidth]{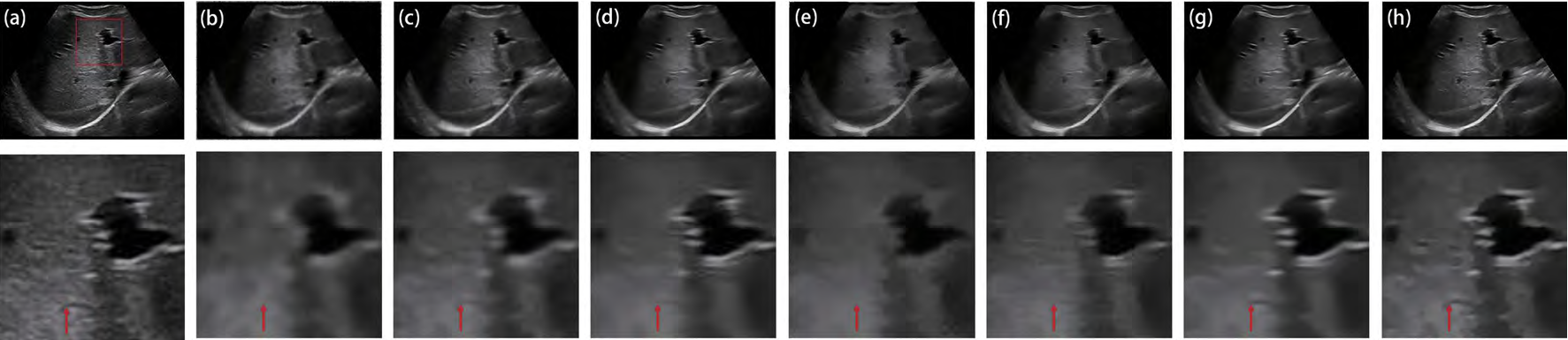}	%调节插图大小
	\caption{Despeckled results of the ultrasound image of liver trauma and  the corresponding zoomed details. (a) The original image; results by (b) Frost, (c) SRAD, (d) OBNLM, (e) SBF, (f) ADLG, (g) NLLRF, and (h) \textbf{PFDTV}.}	%插图名称
	\label{fig6}	%各种label用来引用
\end{figure*}
\begin{figure*}	%插图
	\centering
	\includegraphics[width=\linewidth]{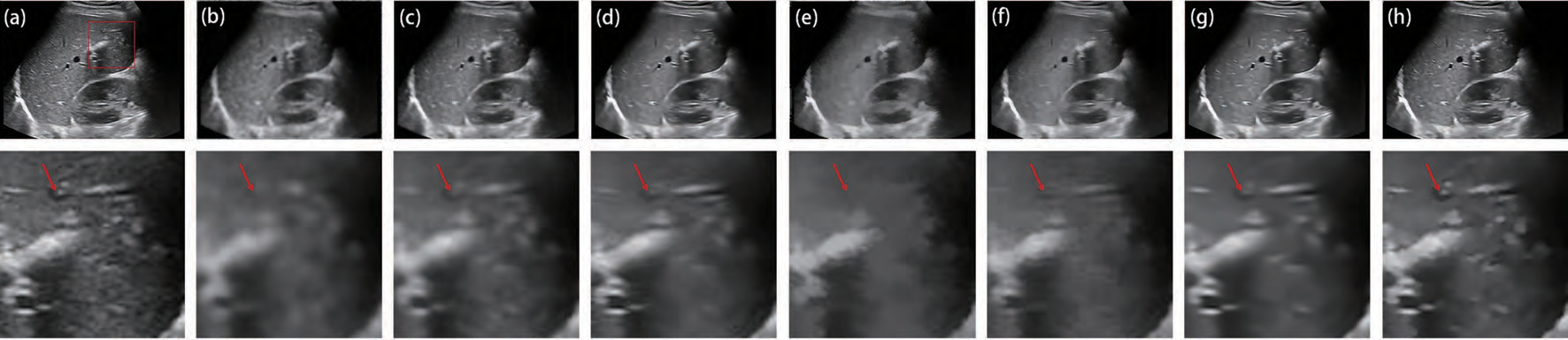}	%调节插图大小
	\caption{Despeckled results of the ultrasound image of hepatitis and  the corresponding zoomed details. (a) The original image; results by (b) Frost, (c) SRAD, (d) OBNLM, (e) SBF, (f) ADLG, (g) NLLRF, and (h) \textbf{PFDTV}. }	%插图名称
	\label{fig7}	%各种label用来引用
\end{figure*}
\begin{figure*}	%插图
	\centering
	\includegraphics[width=\linewidth]{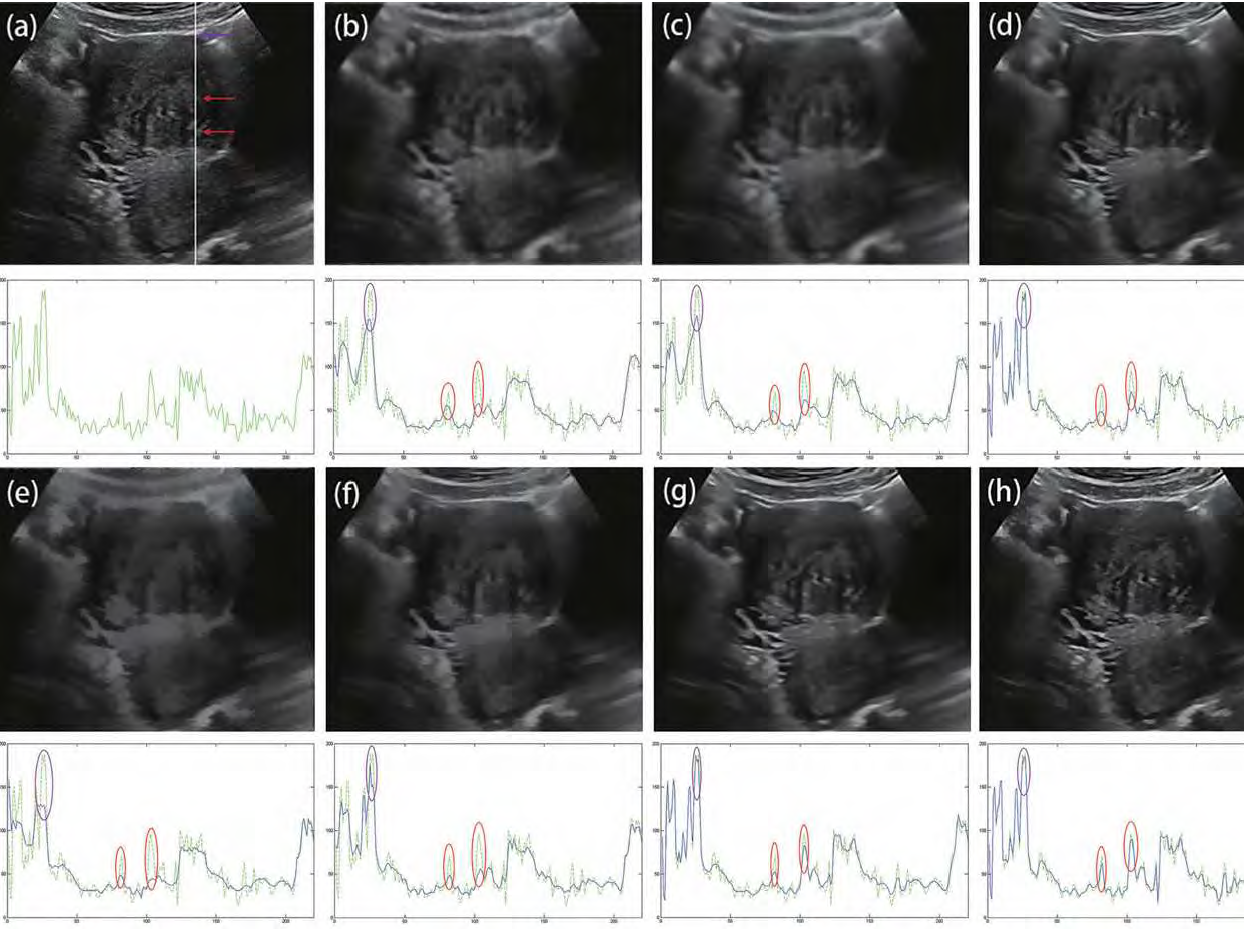}	%调节插图大小
	\caption{Despeckled results of the ultrasound image of uterine fibroids and the corresponding intensity profiles at the same scan line in the white solid line in image (a). The intensity value profile of green line is for the original image while the blue line is for the filtered result of each filter. The purple windows indicate the ROIs for high-contrast edges, while the red windows for low-contrast edges. (a) The original image; the despeckled results by (b) Frost, (c) SRAD, (d) OBNLM, (e) SBF, (f) ADLG, (g) NLLRF, and (h) \textbf{PFDTV}.}	%插图名称
	\label{fig8}	%各种label用来引用
\end{figure*}

The parameter $n_{iter}$ is the iteration number. We integrate $n_{iter}$ into the diffusion coefficient, as shown in (16). If $n_{iter}$ is set to a large number, a small diffusion coefficient will be obtained. This strategy will overcome the problem inherent in the traditional AD model that tends to decrease the edge feature contrast as $n_{iter}$ increases. Fig. \ref{fig5} depicts the despeckled results of different $n_{iter}$. In the case of $n_{iter}=8$, our filter has removed the speckle noise while preserving features properly. As the $n_{iter}$ increases, features are all effectively protected to the same extent. To reduce the required time for despeckling, we set $n_{iter}=8$.

The final optimal parameter configurations of the proposed PFDTV method are set as $\Delta t=0.15,s=15,\mathop k\nolimits_0=20, \mathop n\nolimits_{iter}=8 $. Both the NLLRF and PFDTV methods are employed to perform visual comparison on the same public dataset. Therefore, we set the parameters according to the original paper of NLLRF \cite{20}. Regarding the rest of the filters, we adjust the parameters for each filter to obtain the best visual effect. The final optimal parameter configurations for each filter are set as: 1) Frost: $W = 5 \times 5$, 2) SRAD: $\Delta t = 0.1,\mathop n\nolimits_{iter}  = 120$, 3) OBNLM: $M = 3,\alpha  = 6,h = 1$, 4) SBF: $W=3\times 3,{{n}_{iter}}=15$, 5) ADLG: $\Delta {\rm{t}} = 0.15,\mathop n\nolimits_{iter}  = 80$, and 6) NLLRF: $\beta=10,H=10$. Then, the filters were applied to clinical ultrasound images with their own parameter configurations.

Fig. \ref{fig6} and Fig. \ref{fig7} depict the despeckled results of different filters in the first row and show the corresponding local zoomed-in results in the second row. Visually, our method achieves the best performances in feature preservation and noise removal. According to the streak shown by the red arrow in Fig. \ref{fig6}, the PFDTV method produces the clearest edges. NLLRF only preserves a part of the streak while SRAD reduces the contrast of the streak heavily. Other filters remove the streak after speckle reduction. Similarly, according to the nodules indicated by the red arrow in Fig. \ref{fig7}, the PFDTV method succeeds in enhancing the local contrast. SRAD, OBNLM and NLLRF reduce the contrast of the nodules heavily. Other filters remove the nodules after despeckling.

\begin{figure*}	%插图
	\centering
	\includegraphics[width=\linewidth]{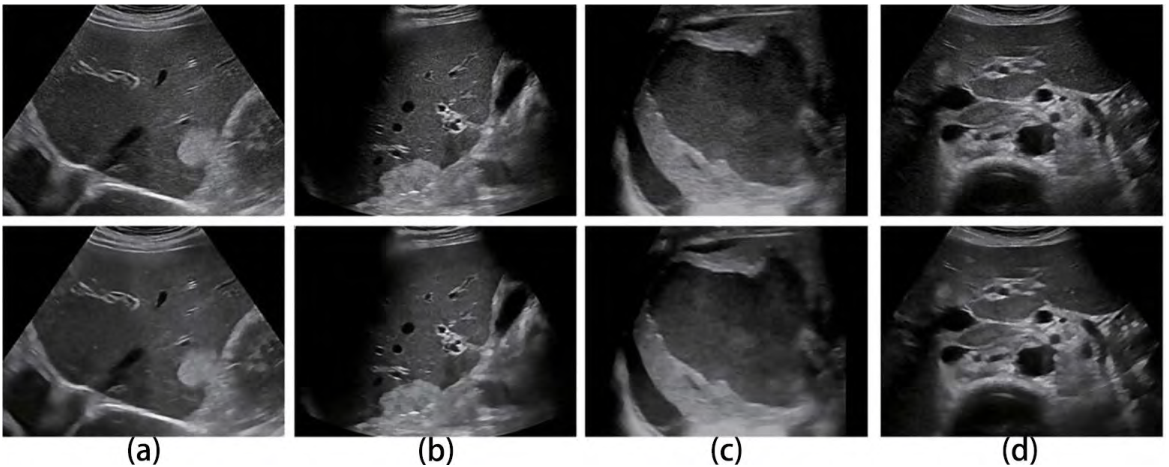}	
	\caption{More despeckled results of the proposed PFDTV method. First row: original ultrasound images; second row: the despeckled results. (a) ultrasound image of haemangiomas, ultrasound (b) image of haemangiomas, (c) ultrasound image of spleen trauma, and (d) ultrasound image of retroperitoneal lymph nodes and tumours image.}	
	\label{fig9}	%各种label用来引用
\end{figure*}

To more closely evaluate despeckled images of different filters, we adopted the method in \cite{69} which evaluates all the features located in a single scan line through the ultrasound image. Fig. \ref{fig8} shows the despeckled images of different filters and their corresponding intensity value profile at the same scan line (marked as white solid line across the image in Fig. \ref{fig8}(a)). The intensity profile of scan line shows that, Frost, SRAD, SBF and ADLG all fail to maintain the edge contrast, reducing the visual effect. OBNLM and NLLRF succeed in enhancing the high-contrast edges in regions of interest (ROI) shown by the purple window. For low-contrast edges indicated by the red windows, both the OBNLM and NLLRF methods fail to preserve the local contrast of edge. As shown by the ROI in red windows in Fig. \ref{fig8}(h), after edge enhancement using adaptive fractional-order $\alpha$, there are slight differences of edge contrast between the despeckled image of the PFDTV method and the original ultrasound image. Compared with other filters, the PFDTV method achieves the best performance in preserving the edge contrast. More despeckled results are depicted in Fig. \ref{fig9}. Obviously, the PFDTV method removes speckle noise thoroughly while preserving features satisfactorily.  

\subsection{Application to ultrasound image segmentation}
\begin{figure*}	%插图
	\centering
	\includegraphics[width=\linewidth]{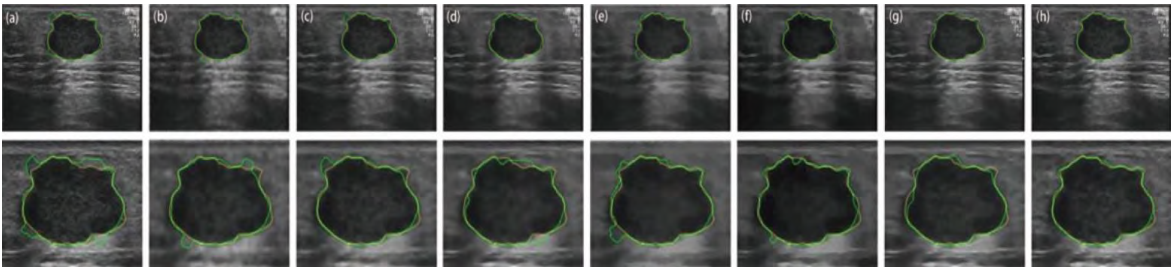}	%调节插图大小
	\caption{The comparison of breast tumour segmentation on different despeckled results. Yellow colour: the ground truth delineated by an experienced clinician; Green colour: the segmentation results produced by . (a) The original ultrasound image and its segmentation result. Despeckled result and the corresponding segmentation results by (b) Frost, (c) SRAD, (d) OBNLM, (e) SBF, (f) ADLG, (g) NLLRF, and (h) \textbf{PFDTV}.  }. 	
	\label{fig10}	%各种label用来引用
\end{figure*}

For validating the performance of our method, we apply each filter to breast ultrasound (BUS) image segmentation. BUS images are commonly used to differentiate between benign and malignant tumours, which can be characterized by their shapes or contours of segmented breast lesions \cite{8}. We first despeckle ten breast ultrasound images with different lesions using different filters. A level-set method\cite{70}\footnote{\url{http://www.imagecomputing.org/~cmli/DRLSE/}} is employed to segment the despeckled results. To validate the segmentation performance for all the ten images, we use the mean value of DSC metrics by averaging the metrics of ten segmentation results to evaluate each filter's despeckling effect on the segmentation performance. Specifically, we first set the range of related parameters of level-set method, then choose the optimal parameters that achieve the highest scores of the mean DSC for the ten BUS segmentation results. Fig. \ref{fig10} displays the BUS segmentation\footnote{The segmentation results of different filters on the ten images are attached in the supplemental file.}: the green curves are the segmentation results of different filters and the yellow curve is the ground truth delineated by an expert; after the speckle reduction, each filter improves the performance of the segmentation result compared with the original BUS image. Among these filters, the segmentation result based on the PFDTV filter is closest to the ground truth. The filtered lesions via other filters are very blurry so that the segmentation method \cite{70} is not able to accurately delineate the lesions from these despeckled images.

We adopt four evaluation metrics \cite{71}, including the dice similarity coefficient (DSC) \cite{40}, Jaccard similarity (JS) \cite{72}, Hausdorff distance (HD) and Hausdorff mean (HM) \cite{73} to measure the segmentation accuracy. As simple spatial overlap indices and producibility validation metrics, DSC and JS measure the overlapping rate between the obtained segmentation region and the ground truth. As effective distance metrics between two finite point sets, HD and HM compute the distance of the contours between the obtained segmentation region and the ground truth. Hence, a better segmentation result should have higher DSC and JS, as well as lower HD and HM. Table \ref{tab2} and Table \ref{tab3} list the mean and median values of DSC, JS, HD and HM for different segmentation results on ten despeckled BUS images, respectively. Obviously, PFDTV achieves the largest DSC and JS values, as well as the smallest HD and HM values, which indicate that PFDTV achieves better segmentation performance compared with other filters.

The computation times for the OBNLM, NLLRF and PFDTV methods are measured and compared on a computer with an Intel(R) Core(TM) CPU at 2.71 GHz and with 8 GB RAM. The computation time of our method on Fig. 7(a) which is a 225$\times$300 image is 86.54 seconds. OBNLM needs 2.85 seconds to obtain the denoised result while NLLRF needs 430.21 seconds. Although PFDTV is not the slowest but its most implementation is C++ code while OBNLM and NLLRF are implemented based on MATLAB, its computation is still inefficient since it needs many more pixels to calculate the fractional-order differential.

\begin{table}
	\centering
	\fontsize{9}{14}\selectfont
	\caption{The Mean DSC, JS, HD and HM values for different segmentation results on ten breast ultrasound images.}
	\label{tab2}
	\begin{tabular}{lcccc }
		\hline
		&DSC(\%)&JS(\%)&HD&HM\cr\hline
		Input&91.87&85.02&16.9933&3.3599\cr
		Frost&93.62&88.02&9.8418&2.1265\cr
		SRAD&94.02&90.52&8.2271&1.5781\cr
		OBNLM&95.64&90.89&8.0608&1.6167\cr
		SBF&93.65&87.66&12.8664&2.0820\cr
		ADLG&94.92&90.37&9.0798&1.7034\cr
		NLLRF&95.39&91.21&7.2801&1.4977\cr
		\textbf{PFDTV}&{\bf96.25}&{\bf92.06}&{\bf4.8441}&{\bf1.2108}\cr\hline
	\end{tabular}
\end{table}
\begin{table}
	\centering
	\fontsize{9}{14}\selectfont
	\caption{The Median DSC, JS, HD and HM values for different segmentation results on ten breast ultrasound images.}
	\label{tab3}
	\begin{tabular}{lcccc }
		\hline
		&DSC(\%)&JS(\%)&HD&HM\cr\hline
		Input&91.63&84.56&16.8859&3.1183\cr
		Frost&93.12&87.12&8.9443&1.9195\cr
		SRAD&95.27&90.97&8.0868&1.5573\cr
		OBNLM&95.45&91.03&6.6623&1.6149\cr
		SBF&94.19&89.02&13.3033&1.9863\cr
		ADLG&94.99&90.49&8.0007&1.6920\cr
		NLLRF&95.5&91.39&7.0000&1.4623\cr
		\textbf{PFDTV}&{\bf96.15}&{\bf92.59}&{\bf4.6926}&{\bf1.3004}\cr\hline
	\end{tabular}
\end{table} 

\section{Discussion and Conclusion}\label{sec5}
Due to the outstanding performance, non-local filtering is becoming a widely accepted method in image restoration and denoising \cite{74}. In ultrasound despeckling, several state-of-the-art non-local filters \cite{6}\cite{9}\cite{10}\cite{18}\cite{20}\cite{75} were developed recently. Compared with other despeckling techniques, they all achieved the best performance of preserving features. To verify the performance of our method, two non-local filters including OBNLM and NLLRF were used for comparison. We find that non-local filters make some low-contrast features heavily blurred. This occurs because the patches around these features are rather similar to the patch centred at speckle noise. After non-local filters remove noise by a weighted average of similar features, these low-contrast features will become indiscernible \cite{39}. Compared with non-local filters, our method performs better in preserving features while removing noise thoroughly.

The most important feature of fractional models is nonlocality. In our increasingly interconnected world, nonlocal interactions are becoming increasingly prominent to spur the studies on nonlocal modeling, analysis and computation. In computer vision and image processing, image denoising \cite{35}\cite{76} can be understood as finding and averaging local and/or nonlocal similar patches for image reconstruction. The nonlocal similarity modeling that considers both geometric and photometric similarity \cite{75}\cite{77} between the similar patches around reference and the selected pixels can be exploited to boost the performance of the feature-preserving ultrasound speckle filtering. In our previous work of texture-preserving nonlocal image denoising \cite{2}, the proposed ACVA method achieves excellent texture-preserving Gaussian and Poisson-Gaussian denoising performance both quantitatively and visually. Thus, an important future direction is to develop a nonlocal \cite{78}\cite{79} version of the PFDTV method for enhancing the texture preservation in ultrasound despeckling. 
	
Image denoising can also be considered as local and nonlocal regression problem \cite{35}\cite{76}\cite{80} that reconstucts the origial signal from local and nonlocal noisy measurments. Most importantly, 3D ultrasound imaging \cite{7}\cite{81} has benefitted a lot from deep learning methods, where deep encoder–decoder architectures serve as a general model for many regression problems and are widely used in computer vision and medical imaging. However, the large flexibility and capacity of deep learning architectures can make them overparameterized for removing the complex speckle noise from ultrasound image. Therefore, how to develop generative adversarial network and/or fractional-order deep network \cite{82} to deal with the statistical characteristic of the signal-dependent speckle noise for improving computational efficiency and accuracy will be interesting research topics in future. 

This paper uses the setting of fixed scale $s$ to simplify our ultrasound despeckling computation without sacrificing significant despeckling performance when the fixed scale being integrated with adaptive fractional order during ultrasound despeckling. The solely setting of fixed scale \cite{39} delivered a less satisfactory performance compared with updating optimal scale for each iteration during ultrasound despeckling. Future work will explore updating scheme with locally adaptive estimation of optimal scale that characterizes the intrinsic local salient structure \cite{38}\cite{39} in ultrasound images.

In conclusion, we have proposed a phase asymmetry guided adaptive fractional-order TV framework combining FAD and FTV filters to achieve feature-preserving ultrasound despeckling.  Synthetic and clinical ultrasound image experiments indicate that the proposed fractional-order TV filter outperforms other well-known ultrasound despeckling filters in both speckle reduction and feature preservation. 

\section*{Acknowledegment}
The authors would like to thank all the cited authors for providing the source codes used in this work and the anonymous reviewers for their valuable comments on the manuscript.

\ifCLASSOPTIONcaptionsoff
  \newpage
\fi



%\bibliographystyle{IEEEtran}



%\newpage
\vfill
Kunqiang Mei received the B.Sc. degree in Biomedical Engineering from Nanjing University of Aeronautics and Astronautics, Nanjing, and the M.Sc. degree in Biomedical Engineering from Shanghai Jiao Tong University, Shanghai, China, in 2016 and 2019, respectively. His current research interests include image processing, machine learning and computer vision.

\vfill
Bin Hu received her MS and MD-PhD degrees from Tongji University and Shanghai Jiao Tong University, Shanghai, China, in 1997 and 2006, respectively. She is a member of the Superficial Organ and Peripheral Vascular Ultrasound Professional Committee of the China Ultrasound Medical Engineering Society and a member of the Ultrasound Branch of the China Association for the Advancement of Medical Sciences. Her current research interest is ultrasound diagnosis and interventional treatment of superficial organ diseases, especially breast diseases.

\vfill
Baowei Fei is a professor in the Erik Jonsson School of Engineering and Computer Science, University of Texas at Dallas, USA. He received his MS and PhD degrees from Case Western Reserve University, Cleveland, Ohio. He is a director of the Quantitative Bioimaging Laboratory (https://fei-lab.org/) at Departments of Bioengineering and Computer Science in University of Texas at Dallas. 

\vfill

Binjie Qin (M’07) received his MS and PhD degrees from Nanjing University of Science and Technology, Nanjing, and Shanghai Jiao Tong University, Shanghai, China, in 1999 and 2002, respectively. He was a lecturer and then associate professor at School of Life Sciences and Biotechnology in Shanghai Jiao Tong University. From 2012 to 2013, He was a visiting professor at Department of Computer Science, University College London, U.K. He is currently an associate professor at School of Biomedical Engineering in Shanghai Jiao Tong University. His current research interests include biomedical imaging, image processing, machine learning, computer vision and biomedical instrumentation.


%\vfill


\end{document}